%% file: main.tex
\pdfoutput=1

\documentclass[11pt]{article}

\usepackage[final]{acl}

\usepackage{times}
\usepackage{latexsym}
\usepackage{enumitem}
\usepackage[T1]{fontenc}

\usepackage[utf8]{inputenc}

\usepackage{microtype}

\usepackage{inconsolata}
\usepackage{subcaption}

\usepackage{graphicx}
\usepackage{ai-usage-card}
\usepackage{annotation}

\usepackage{float}
\usepackage{hyperref}
\usepackage{cleveref}

\usepackage[skip=2pt plus1pt, indent=10pt]{parskip}

\newcommand{\methodName}{CiteAssist}

\title{\methodName: A System for Automated Preprint Citation and BibTeX Generation}

\author{Lars Benedikt Kaesberg \\
  Georg-August University / Göttingen \\
  \texttt{l.kaesberg@stud.uni-goettingen.de} \\\\
  \textbf{Jan Philip Wahle} \\
  Georg-August University / Göttingen \\
  \texttt{wahle@uni-goettingen.de}\\\And
  Terry Ruas \\
  Georg-August University / Göttingen \\
  \texttt{ruas@uni-goettingen.de}\\\\
  \textbf{Bela Gipp} \\
  Georg-August University / Göttingen \\
  \texttt{gipp@uni-goettingen.de}
  }

\begin{document}

\maketitle
\AddAnnotationRef

\begin{abstract}
We present \textit{\methodName}, a system to automate the generation of BibTeX entries for preprints, streamlining the process of bibliographic annotation. 
Our system extracts metadata, such as author names, titles, publication dates, and keywords, to create standardized annotations within the document. 
\methodName{} automatically attaches the BibTeX citation to the end of a PDF and links it on the first page of the document so other researchers gain immediate access to the correct citation of the article (see \autoref{fig:result-overview-system}).
This method promotes platform flexibility by ensuring that annotations remain accessible regardless of the repository used to publish or access the preprint. 
The annotations remain available even if the preprint is viewed externally to \methodName.
Additionally, the system adds relevant related papers based on extracted keywords to the preprint, providing researchers with additional publications besides those in related work for further reading. 
Researchers can enhance their preprints organization and reference management workflows through a free and publicly available web interface\footnote{\url{https://preprint.larskaesberg.de}}. 
\end{abstract}

\section{Introduction}

A common challenge in academic research is tracing citational information for preprints. 
Many researchers experience a low citational impact because their citational information is not accessible easily together with the preprint.
This phenomenon is often evident on platforms like ResearchGate, where numerous articles are downloaded hundreds of times but receive only a few citations \cite{bollmann-elliott-2020-forgetting}. 
This discrepancy between reads and citations highlights a difference in how information is consumed and credited (a pattern also observed in contexts such as social media impressions versus likes) and specifically in citation analysis \cite{meho2006rise,WahleRAG23,wahle2024citation}.

\begin{figure}
\centering
\begin{subfigure}[b]{0.48\textwidth}
   \includegraphics[width=0.95\linewidth]{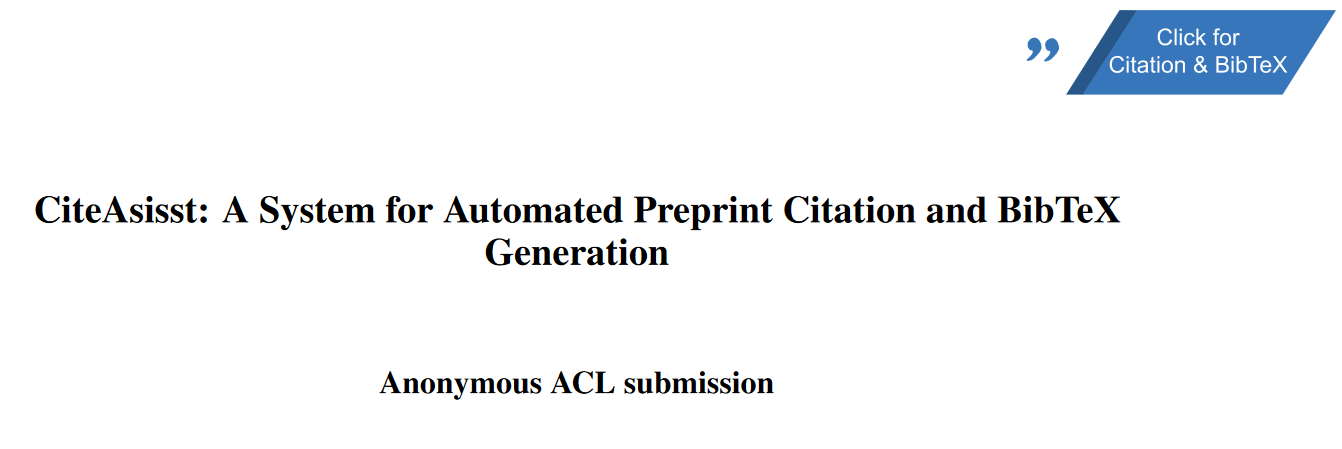}
   \caption{The first preprint page with a generated annotation button.}
   \label{fig:Ng1} 
\end{subfigure}

\begin{subfigure}[b]{0.48\textwidth}
   \includegraphics[width=0.95\linewidth]{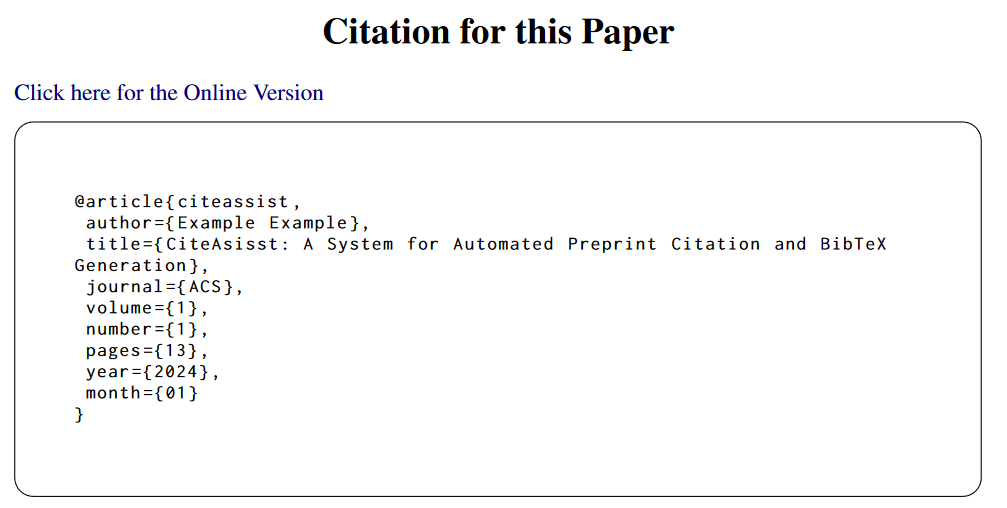}
   \caption{A generated citation box on the last page of the preprint.}
   \label{fig:Ng2}
\end{subfigure}

\caption{Document enhanced with \methodName. A detailed version available in \autoref{sec:enhanced-doc}).} \label{fig:result-overview-system}
\end{figure}
An imbalance between reading and citing papers can have many reasons: researchers may judge a focal work as irrelevant to them, not exciting, not sound, or simply forgetting.
Yet, a key reason is the accessibility of citational information \cite{bollmann-elliott-2020-forgetting}. 
If citation details are consistent and easily accessible, citations are generally higher \cite{citationAdvantage, bollmann-elliott-2020-forgetting}.
Long review processes and publication timelines can slow scientific progress and hinder researchers from proposing state-of-the-art techniques.
Thus, preprints are often used in modern scientific research.
Public available preprints promote the rapid exchange of ideas and results before peer-reviewed publications are available, accelerating the spread of knowledge and driving scientific progress \cite{Puebla2022}. 
With the fast growth in the number of preprints, researchers are faced with an ever-growing sea of scientific information \cite{10.7554/eLife.45133}. 
The sheer volume of preprints make efficient retrieval challenging.

So far, there is no easy way to annotate preprints with citational information. 
While some repositories allow for archiving preprints (e.g., Zenodo, figshare, arXiv), the citation details of a paper are often only available through external infrastructure (e.g., publishers, scholarly search engines).
To the best of our knowledge, no publicly available system provides the citation details of a paper within the PDF itself in simple format.
Furthermore, we lack a common way to add recommended literature on related projects within preprints for further reading about a topic over the scope of related work sections. 
Those could be related ideas from other areas, books, or talks that complement a paper but are not directly discussed in related work.

We present \methodName, a system designed to simplify the creation and distribution of citation annotations by automatically generating and appending annotations to the preprint. 
This is achieved either by generating a new PDF with the annotation or by generating a \LaTeX{} file that can be integrated directly into the source document.
\methodName{} provides researchers with a convenient way to enhance their preprint by directly attaching citation information to it, improving the reference management workflow.
These annotations capture metadata such as author names, publication details, and abstracts, enabling researchers to create bibliographic records for their preprints.
The system also allows users to add custom metadata (e.g., keywords) to the BibTeX annotation.
By automating this process, researchers can save time that would otherwise be spent on manual annotation. 
Our system is platform-independent as it incorporates all citation details within the paper.
The annotation is attached to the preprint; thus, it can be published at any preprint repository.
Unlike platform-specific solutions, our system accommodates various sources, making it accessible and useful to researchers in various fields. 
\autoref{fig:result-overview-system} (a) shows the first page containing the button to the BibTex entry and (b) the generated annotation itself.
In addition, \methodName{} suggests related papers at the end of an article to complement a traditional ``related work'' section. 
Related work sections focus on considering the proposed work and crediting previous ideas. 
However, they are bound to the researcher's awareness of these works, which may include \textit{citation amnesia}, i.e., not citing good and relevant work \cite{singh-etal-2023-forgotten, wahle2024citation}. 
\methodName{} mitigates this issue by suggesting additional literature on potentially forgotten or overlooked works.
Researchers can manually add literature they wish to suggest to readers as additional resources.

\section{Related Work}

\subsection{Preprint Platforms}

Multiple platforms allow researchers to publish their preprints to repositories, e.g., arXiv, bioRxiv, ResearchGate, etc \cite{metaresearchbio,Puebla2022}. 
While these platforms support scholarly communication, they rely on external infrastructure that may not be accessible or familiar to all researchers. 
Some may prefer to upload their work to personal websites or may not know how to use \LaTeX.
Our system bridges these gaps, making preprint submission and formatting easier by providing a simplified interface that allows researchers to enter citation details directly into the paper. 

\subsection{Annotation Generation Tools}
Annotation generation tools automatically extract metadata from scholarly articles and generate annotations. 
In this section, we discuss three relevant annotation generation tools relevant to \methodName{}, namely CERMINE \cite{cermine}, Neural ParsCit \cite{animesh2018neuralparscit}, and Grobid \cite{GROBID}.

\noindent
\textbf{CERMINE} is a tool for extracting metadata from scholarly articles. 
It employs machine learning techniques like SVM, CRF, and clustering to parse PDF documents and extract structured information, including authors, titles, abstracts, and references \cite{cermine}. 
CERMINE can extract metadata from diverse scholarly articles, even with complex layouts and formatting. 
It generates structured XML output, creating annotations in various formats, such as BibTeX or CSL \cite{cermine}.
Unlike CERMINE, which primarily extracts metadata for structured XML output, our system generates BibTeX annotations for scientific citation and proactively suggests related papers, increasing the visibility of preprints in the scientific literature.
\methodName{} integrates related literature, which enhances the annotation process and the researcher's ability to explore related works. 
Additionally, while CERMINE outputs structured XML, our system directly attaches the generated BibTeX annotation to the paper, providing a more streamlined workflow for researchers. 

\noindent
\textbf{Neural ParsCit} focuses on extracting citation information from scientific articles. 
While its primary purpose is to parse and analyze citations, Neural ParsCit can also generate annotations using the extracted information. 
It identifies and extracts key metadata elements, including authors, titles, and publication venues. 
The tool focuses on citations, ensuring accurate extraction of citation-related metadata (e.g., identifying a citation string such as ``Doe, J., Smith, A., \& Johnson, K. (2020). Title of the article. 10(2), 123-134'').
However, Neural ParsCit does not capture information like the publication date or journal \cite{cermine}. 
Therefore, additional processing or integration with other tools may be required to generate annotations for preprints \cite{animesh2018neuralparscit}.
In contrast, our tool provides a complete solution for BibTeX annotation generation and the integration of related papers. 
Unlike Neural ParsCit, \methodName{} leverages user input and keyword-based searches to generate BibTeX citations that are directly linked in the paper PDF and to include relevant literature.

\noindent
\textbf{Grobid} provides an API for extracting metadata from papers, using both a small CRF model for quick results and a larger BidLSTM-CRF model for higher accuracy, though at greater computational cost \cite{GROBID}. 
While it shares capabilities with CERMINE, like extracting titles, authors, and abstracts, Grobid does not capture journal information \cite{GROBID}. 
Although Grobid’s self-hostable API via Docker makes it an excellent component for \methodName{} to leverage for more advanced data extraction directly from the PDF text, it lacks the advanced PDF enhancement features that \methodName{} offers, such as direct annotation with BibTeX and the suggestion of related papers.

\section{System Architecture}
The system architecture of \methodName{} consists of a frontend, backend, and database. 
The frontend handles the user interaction and allows users to upload a PDF file, which is parsed by our method and in the backend with Grobid. 
We merge and display the results to the user for correction, if necessary.
Furthermore, users can add related papers, either provided by the backend via keyword matching or with a provided DOI or arXiv id. 
The system generates a new PDF or \LaTeX{} file with an attached BibTeX annotation and related papers. 
After user approval, these results are uploaded to the backend, where the PDF is stored locally and metadata forwarded to the database.
We refer to \autoref{sec:system-architecture} for a more detailed system architecture description.

\section{Preprint Annotation Workflow}

This section shows a practical example of how researchers can annotate preprints using \methodName. 
The workflow begins when a user uploads a preprint in PDF format to the \methodName{} homepage via a drag-and-drop interface or a file selection dialogue as seen in \autoref{fig:homepage}. 
Users are then asked to review (\autoref{fig:metadata}), edit potentially incorrect information, or add other relevant metadata (e.g., keywords).
Users can select relevant suggested keywords and add new ones if desired. 
The system enhances the compatibility of these keywords through lemmatization. 
With refined keywords, the backend searches the database for related preprints and displays these on the frontend for user review and selection, as seen in \autoref{fig:related}. 
\methodName{} compiles this information and generates a BibTeX annotation containing the edited metadata and selected related papers (\autoref{fig:annotation-large}). 
The compiled annotation is attached at the end of the preprint.
On the title page of the uploaded article, it adds a button that links to the annotation at the bottom, as shown in \autoref{fig:button}.
Users can download the final annotated PDF version or generate \LaTeX{} files to include it directly in the source document. 
With user consent, the backend stores the annotated PDFs for future access. 
Users can view these files online, share them with colleagues, or download them again.

\section{BibTeX Annotation Generation}

We combine the \texttt{PDF-LIB} \footnote{\url{https://github.com/Hopding/pdf-lib}} library, Grobid, and our custom information extraction method to extract BibTeX annotation information. 
Our process can be divided into five steps:

\begin{enumerate}[wide, labelwidth=!, labelindent=0pt]
    \item \textbf{PDF Parsing:} The PDF file is parsed using the \texttt{PDF-LIB} library, which allows us to access the document's metadata (e.g., title, author, creation date) if available and saved in the PDF.
    
    \item \textbf{Text Extraction:} Our custom function extracts the text content from the PDF file. This includes the text from the first page, which contains information such as the title, author, and more. The rest of the text is used for keyword extraction.
    
    \item \textbf{Metadata Extraction:} We extract various metadata elements necessary for the BibTeX annotation. These elements include:
    
        \begin{itemize}
            \item \textbf{Article Type:} The publication category for the paper (e.g., proceedings, journal). Set to “article” as a default.
            \item \textbf{Author:} Author name obtained from the PDF metadata or extracted from the text.
            \item \textbf{Title:} Title of the article from the first page's text. If unavailable, the PDF's title or the filename is used as a default.
            \item \textbf{Pages:} The total number of pages in the PDF document.
            \item \textbf{Date:} Publication date from the PDF metadata. If unavailable, the current date is used.
        \end{itemize}
    \item \textbf{Grobid:} Simultaneously, the frontend sends the PDF to the backend, which then forwards it to Grobid to extract relevant information, including the title, author, publication date, and keywords. %
    \item \textbf{Constructing the BibTeX Annotation:} We merge the information extracted from  Grobid and our own. In cases where neither alternative provides the information, we resort to a fallback value that can be edited by the user. Next, we construct the BibTeX annotation following the required format from the extracted metadata. The annotation includes key-value pairs for elements such as author, title, journal, volume, pages, year, and DOI.
    
\end{enumerate}

The combination of \texttt{PDF-LIB}s metadata extraction, Grobid, and our custom metadata extraction function allows us to generate BibTeX annotations for preprints shown in \Cref{fig:Ng2} (a figure outlining the process can be found in \autoref{fig:annotation-process}).

\section{Integration of Related Papers}
The integration of related papers is designed to augment the research experience by providing users with access to literature that, while not directly cited as related work, can indicate the reader to other possible related materials.
We use keyword-based searches to retrieve a selection of papers that serve as complementary references, offering broader insights into the subject area without overlapping with the core related work already cited.
This task involves the following functionalities:

\begin{itemize}
    \item \textbf{Keyword Extraction:} The preprint's whole text is processed by our algorithm to extract the five most relevant keywords. This is done by removing stop words, applying lemmatization, and selecting based on frequency of occurrence. This functionality is currently only available for English texts. If the author has provided keywords, Grobid will also extract these. 
    
    \item \textbf{User Input:} Users provide additional keywords or tags associated with the preprint being annotated. These keywords act as search criteria for finding related papers.
    
    \item \textbf{Search Query:} The system applies lemmatization to the provided keywords, converting them to their base forms for better matching. This improves the effectiveness of the query in retrieving related papers. The query returns the five most relevant papers, ranked by the number of matching keywords.
    
    \item \textbf{Selection and Integration:} Users can choose specific preprints from the provided suggestions to add to their annotation. They can also add new papers using either a DOI or arXiv ID. These selected papers are integrated into the annotation generation process.

\end{itemize}

Integrating related papers into the annotation enhances the paper by providing users access to related resources that contribute insights, broader the existing related work section and further context to their initial findings, and save researchers time and effort from authors and readers.

\section{Final Considerations}

In this work, we introduced \methodName, a system designed to streamline the creation and distribution of BibTeX annotations within preprints. 
\methodName{} standardizes the reference management process with the help of automatic generation and integration of annotations, either through a newly generated PDF or directly within a \LaTeX{} source document.
A button on the title page linking to the annotation improves the usability and accessibility of the citation information of the paper.
Our system enhances the usability of preprints by embedding citation details and ensures that the annotations are readily accessible, regardless of the publication repository used. 
Additionally, \methodName{} offers suggestions for related papers, enriching the scope of resources available to researchers about a given publication.
\methodName's development reflects the growing importance of digital tools in academic research. 
By allowing researchers to easily include their citational information into preprints, we see a direct impact on giving authors proper credit and to be cited correctly with the publication venue as opposed to the preprint server. 
Future enhancements could include recommending similar papers based on semantic similarity and providing a more comprehensive representation of the paper's content. 

\section*{Limitations}

Although \methodName{} has enhanced the preprint annotation process, it still has room for improvement.
Implementing stopword removal for several languages during keyword extraction may improve the relevancy of the discovered keywords.
Furthermore, integrating language models to extract keywords directly can take advantage of recent natural language processing advances (e.g., large context length \cite{ma2024megalodon}) to produce more accurate results.
Incorporating large language models into \methodName{} can improve metadata extraction accuracy and cover a wider range of document layouts and formats, increasing the tool's overall efficiency. 
We plan to improve \methodName{}'s UI and usability through user feedback sessions. 
By gathering insights directly from users, we will make iterative improvements to enhance the system’s intuitiveness and functionality, ensuring it effectively meets researchers' needs.

\bibliography{rebiber_sources}
\appendix
\section{Detailed System Architecture}
\label{sec:system-architecture}

The system architecture of \methodName{} is designed to provide researchers with a responsive and user-friendly experience. 
It consists of a frontend, a backend, and a database, each component fulfilling specific functionalities to enable efficient and semi-automatic annotation generation.

\subsection{Frontend and UI/UX Design}

\begin{figure*}[ht]
    \centering
    \includegraphics[width=0.8\textwidth]{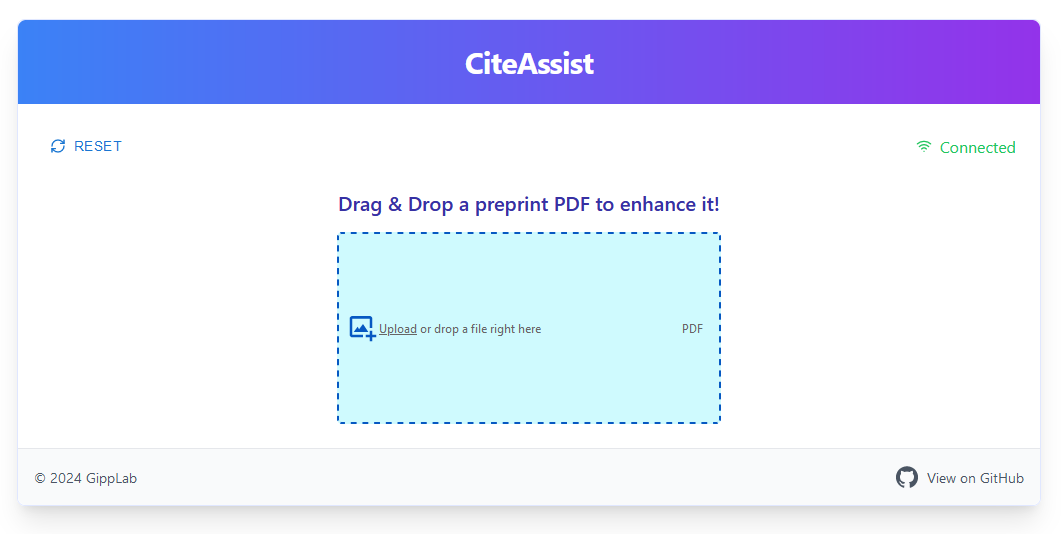}
  \caption{\methodName{} Homepage.}
    \label{fig:homepage}
\end{figure*}

\begin{figure*}[ht]
    \centering
    \includegraphics[width=0.8\textwidth]{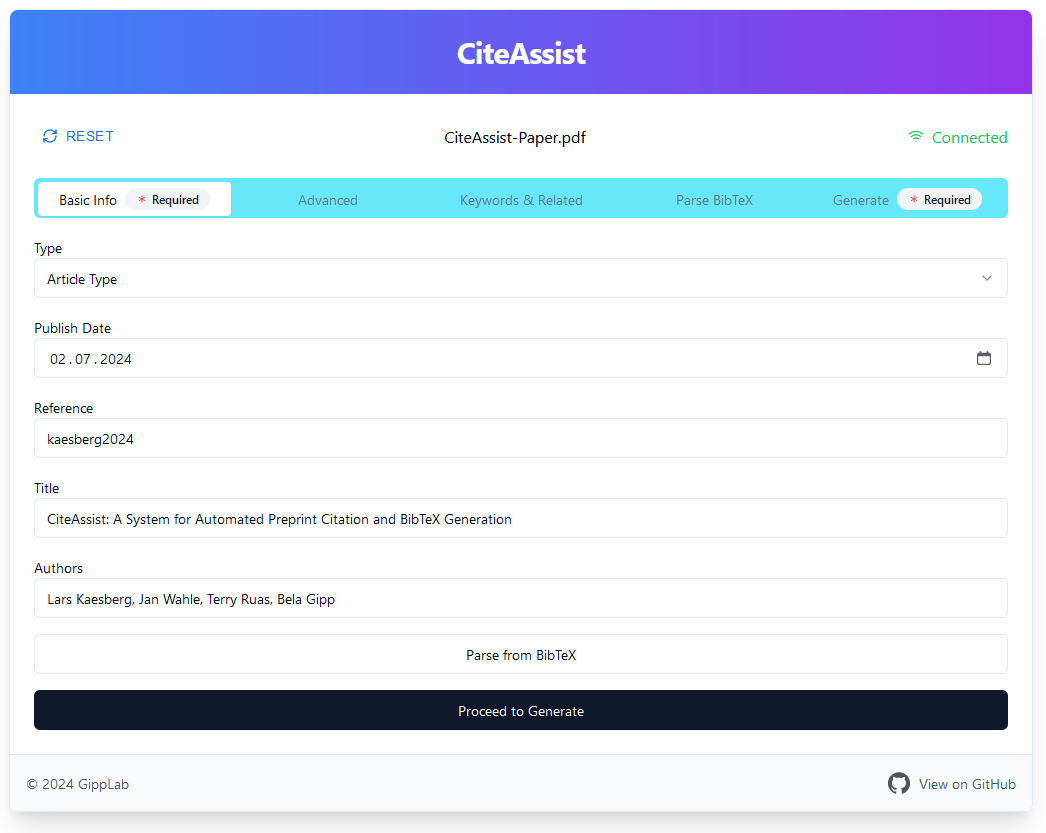}
  \caption{\methodName{} Edit Metadata Page.}
  \label{fig:metadata}
\end{figure*}

Our system's front end handles the user's interaction with the system. 
Researchers can upload their preprint in PDF format, and the system extracts metadata such as the preprint's title, author, year, and page numbers to generate the BibTeX annotation (see \autoref{fig:homepage}). 
This process is done both in the frontend and in the back end with the help of Grobid, and then results are merged to achieve the most complete set of automatically extracted metadata.
The data is displayed to the user, offering an overview of the preprint's details (see \autoref{fig:metadata}). 
To guarantee accuracy, users can edit this extracted data as needed. 
Also, users can directly copy and paste a BibTeX they already have (for example from their venue) and the tool extracts the necessary metadata from that BibTeX.
Once verified, the frontend sends the metadata, along with the user-specified keywords, to the backend.

Beyond metadata extraction, the frontend is responsible for generating the final PDF file. It receives related papers from the backend and integrates them with the BibTeX annotation into a new PDF document. The resulting PDF combines the preprint with the annotation and related literature. The system can also create a \LaTeX{} file with the annotation that can be imported into the \LaTeX{} source document. This makes it convenient to submit to source code-based repositories such as arXiv.

\noindent
\textbf{Design Principles:}
We dedicated our UI/UX design to user-centric principles. To support our design decisions, we looked at reputable UI/UX guidelines, such as those proposed by \citet{usability}. Adopting these universally acknowledged design principles ensures our tool meets industry standards. We prioritized:
\begin{itemize}
    \item \textbf{Simplicity:} We aimed for a clean and uncluttered interface, prioritizing simplicity to ensure ease of use. To achieve this, we employed a minimalist design approach, using a monochromatic color scheme and limiting the number of interactive elements visible at any one time. This approach reduces cognitive load and allows users to focus on task-relevant information.
    
    \item \textbf{Consistency:} We maintained consistent visual and interaction design throughout the tool to provide a cohesive user experience. We adopted a standard color palette and uniform typography across all pages. Navigation elements are consistently placed in the same locations across screens.
    
    \item \textbf{Intuitiveness:} We focused on creating an intuitive user interface, where users can easily understand the tool's functionalities without extensive instructions. Labels are descriptive and positioned adjacent to their respective controls, tooltips are provided for complex functions.
    
    \item \textbf{Responsive Design:} We adopted a responsive design approach, ensuring that the UI adapts seamlessly to different screen sizes and devices. We used a flexible grid layout to adjust the layout dynamically, providing an optimal viewing experience on tablets, phones, and desktops.
\end{itemize}

\begin{figure*}[ht]
    \centering
    \includegraphics[width=0.8\textwidth]{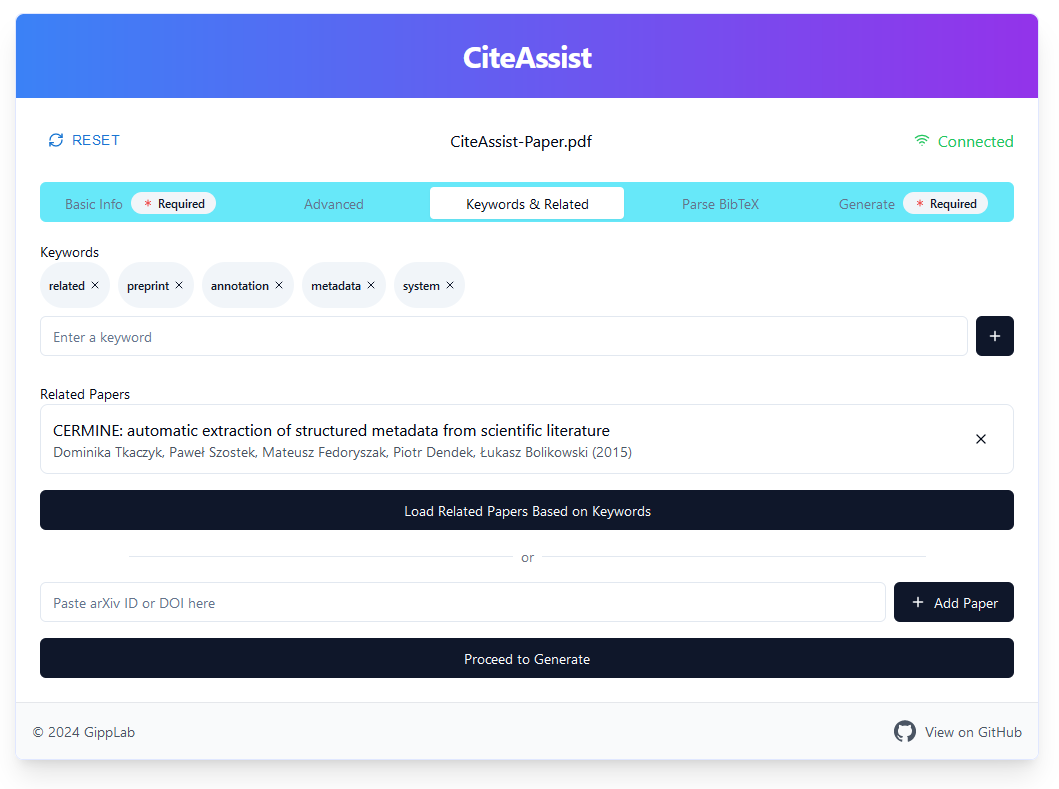}
  \caption{\methodName{} Related Paper Page.}
  \label{fig:related}
\end{figure*}

\noindent
\textbf{User Interface Components:}
The user interface components are designed to address user tasks through distinct components.
\begin{itemize}
    \item \textbf{Upload Interface (\autoref{fig:homepage}):} Users are presented with an interface to upload their preprint PDF files. The upload interface includes drag-and-drop functionality and instructions to guide users through the process.
    
    \item \textbf{Metadata Editing (\autoref{fig:metadata}):} Users have the ability to edit the extracted metadata of the preprint, such as title, author, etc. A user-friendly form-based interface is employed, allowing users to make necessary modifications easily.
    
    \item \textbf{Keyword Input (\autoref{fig:metadata}):} Users can input relevant keywords or tags associated with the preprint to search for related papers. The keyword input interface supports both manual entry and suggestions based on extracted keywords.
    
    \item \textbf{Related Papers Display (\autoref{fig:related}):} The interface presents related papers retrieved based on the user's keywords. Users have the option to delete papers they consider irrelevant, or to add related papers with their Digital Object Identifier (DOI) or arXiv id.
\end{itemize}

\noindent
\textbf{User Experience Enhancements:}
We took steps to enhance user interaction with our system. This included:
\begin{itemize}
    \item \textbf{Ease of Navigation:} The system provides seamless navigation between its different sections. To enhance ease of navigation, clear and intuitive navigation elements, such as a persistent top menu, have been implemented.
    
    \item \textbf{Feedback and Error Handling:} The tool provides error messages to guide users in case of incorrect inputs.  Clear error messages, tooltips, and validation checks have been implemented to assist users in understanding and resolving issues effectively.
    
    \item \textbf{Visual Appeal:} The UI utilizes visually appealing design elements, such as appropriate color schemes, typography, and layout, to provide an engaging and aesthetically pleasing experience for users.
\end{itemize}

\subsection{Backend}
The backend component of our system searches for related preprints based on provided keywords.
The primary role of the backend is to accept new preprint metadata from the frontend and store them in the database for future reference. 
It acts as a bridge, receiving the preprint information and passing it on to the database without directly processing the preprint content.

In addition, the backend supports querying the database of previously uploaded preprints for related works based on keywords provided by the user. 
When the frontend sends a request for related preprints, the backend performs the necessary database searches to identify relevant preprints.

The backend component of our tool also stores the actual PDF files of the preprints. 
This storage capability is crucial, as it allows for an easy retrieval of the original documents. 
When a user requests to view or download a specific preprint, the backend serves the corresponding PDF file back to the frontend. 

\subsection{Database}
The database component of our system is responsible for storing and managing information about preprints. 
It serves as a central repository for preprint metadata, enabling efficient data storage and retrieval. 
The database stores information about each preprint, including its title, author, DOI, URL to the publication, year of publication, and keywords associated with it. 

To ensure secure and reliable data storage, our system uses PostgreSQL as the database management system. PostgreSQL offers robust features such as data integrity, efficient indexing, and transaction support, making it well-suited for managing preprint information.
\onecolumn
\section{\methodName{} Screenshots}

\begin{figure*}[ht]
    \centering
    \includegraphics[width=0.8\textwidth]{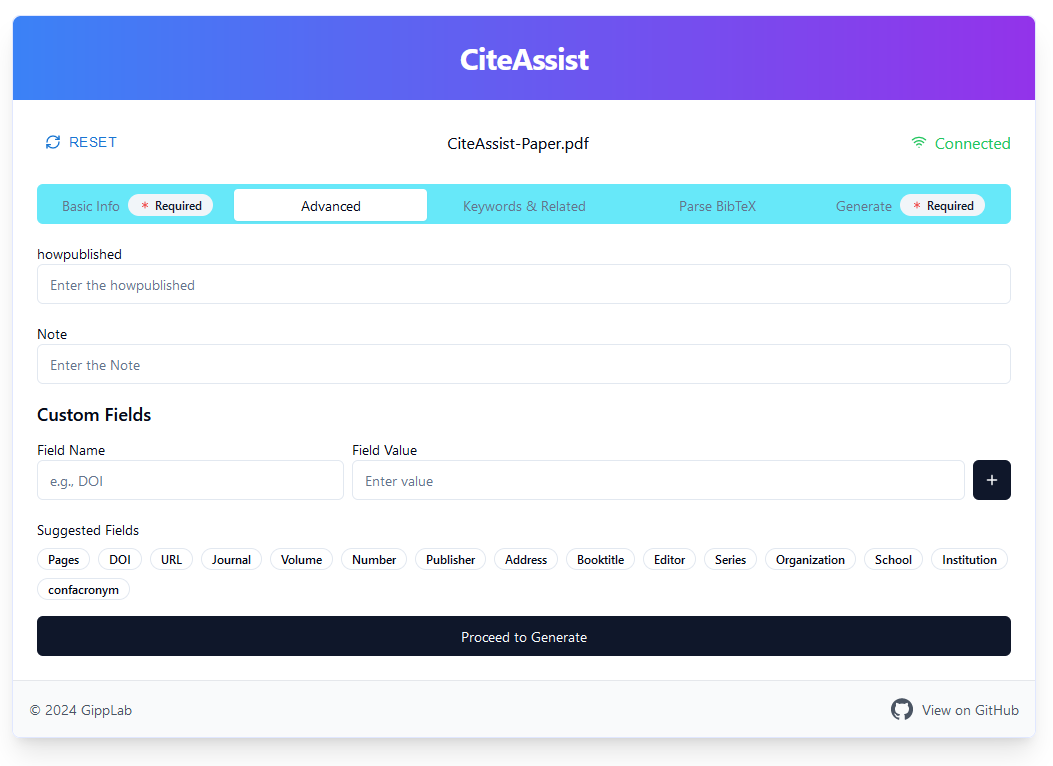}
  \caption{\methodName{} Advanced Edit Metadata Page.}
\end{figure*}

\begin{figure*}[ht]
    \centering
    \includegraphics[width=0.8\textwidth]{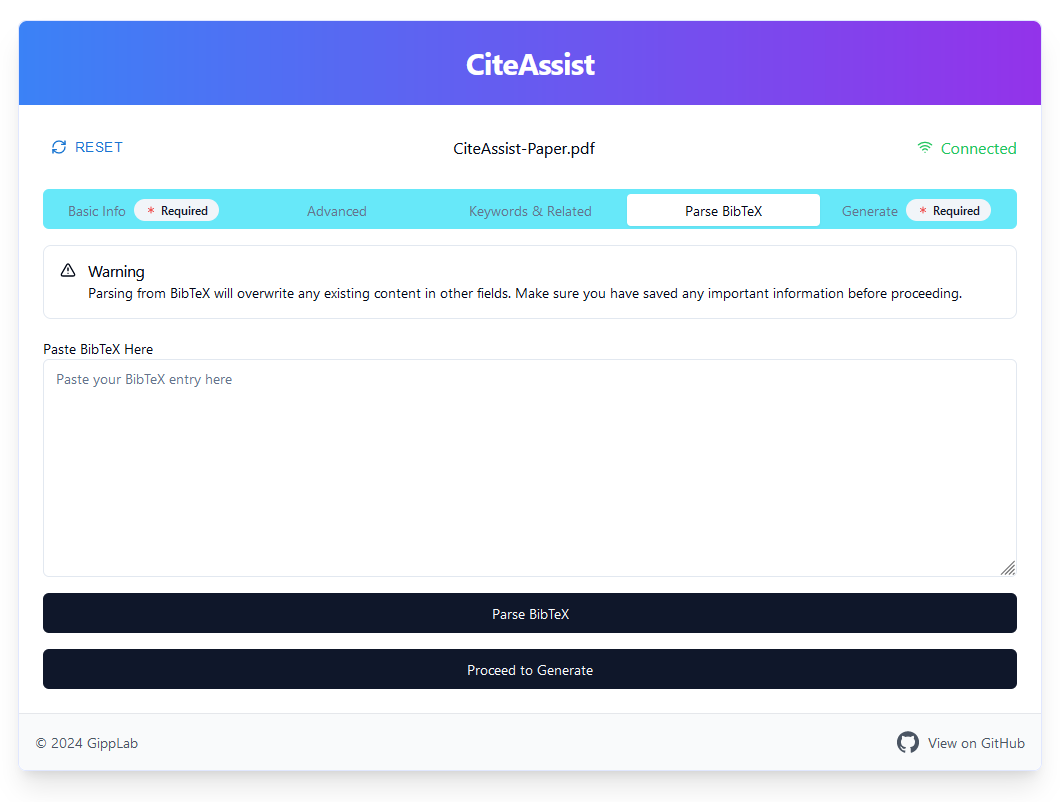}
  \caption{\methodName{} Parse BibTeX Page.}
\end{figure*}

\begin{figure*}[ht]
    \centering
    \includegraphics[width=0.8\textwidth]{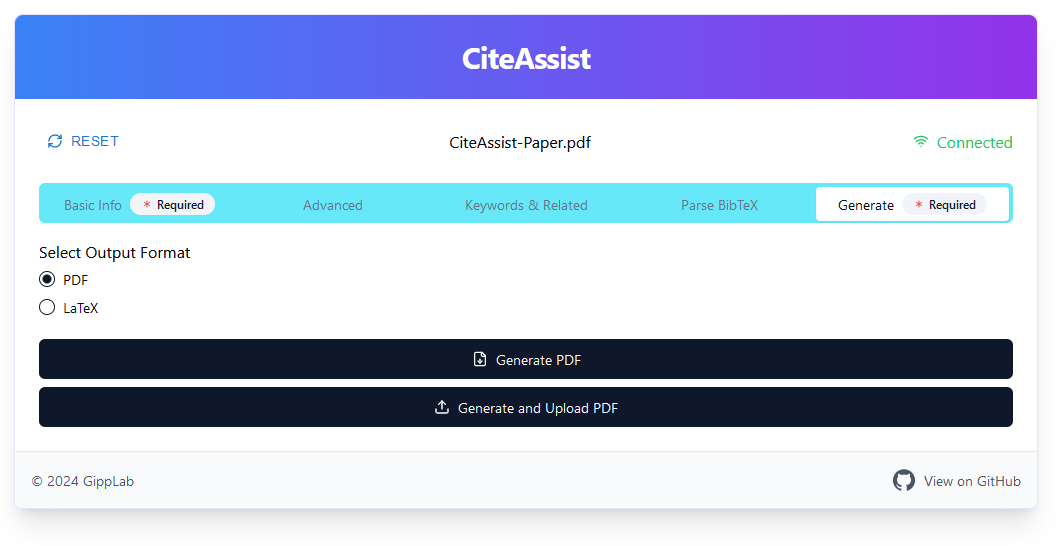}
  \caption{\methodName{} Generate Page.}
\end{figure*}

\begin{figure*}[ht]
    \centering
    \includegraphics[width=0.8\textwidth]{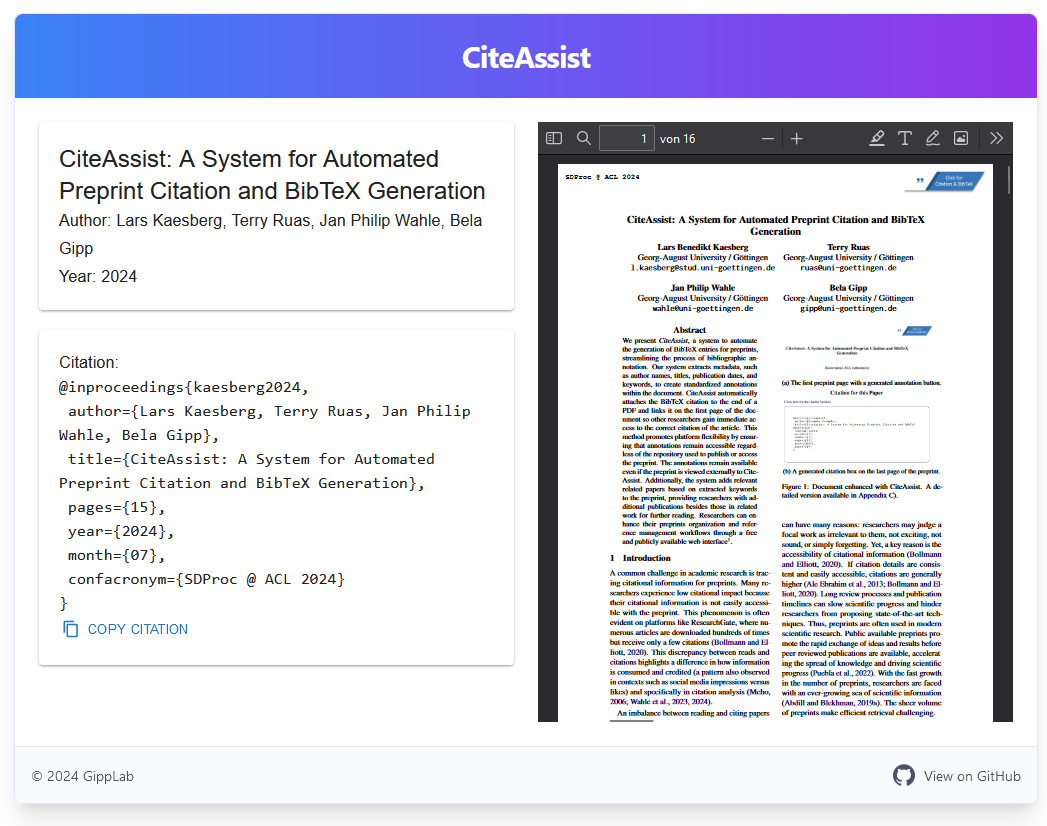}
  \caption{\methodName{} Preprint Webview.}
  \label{fig:webview}
\end{figure*}

\clearpage
\section{\methodName{} Enhanced Document}
\label{sec:enhanced-doc}

\begin{figure*}[ht]
  \centering
    \includegraphics[width=1\textwidth]{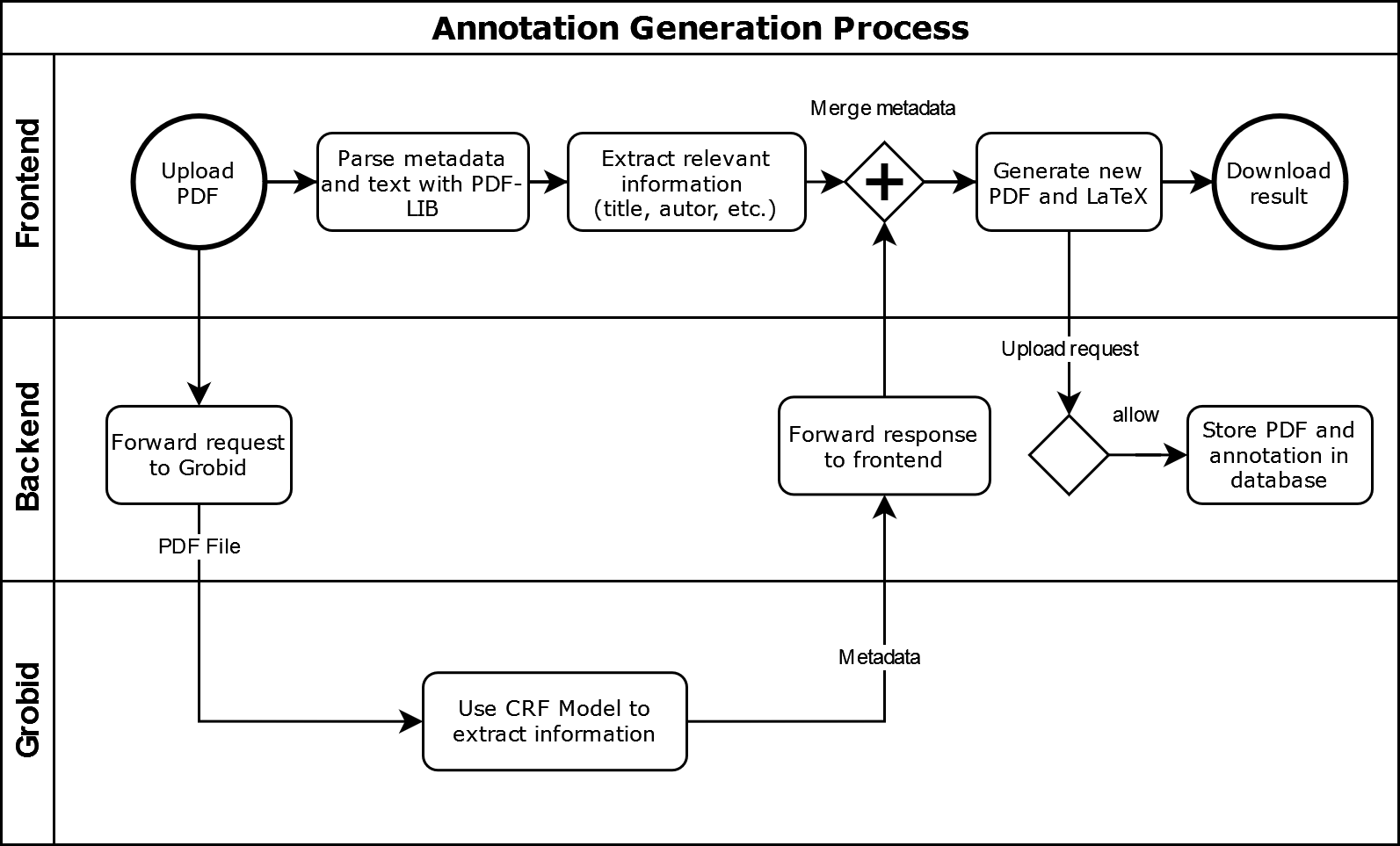}
  \caption{Generation process for a \methodName{} annotation.}
  \label{fig:annotation-process}
\end{figure*}

\begin{figure*}[ht]
  \centering
  \begin{tikzpicture}
    \node[draw,line width=1pt,inner sep=0pt] (image) {\includegraphics[width=1\textwidth]{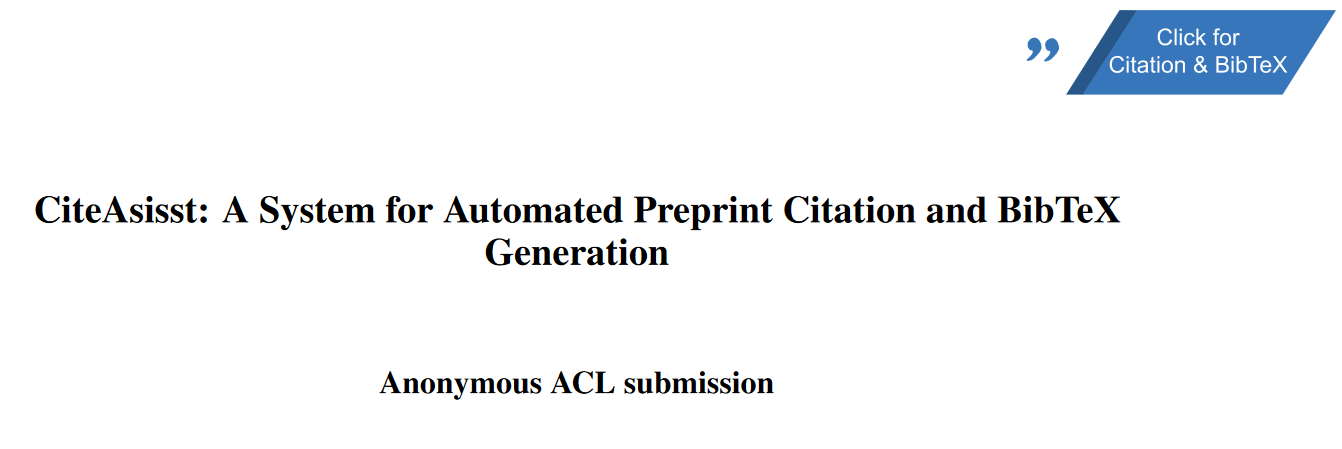}};
  \end{tikzpicture}
  \caption{The first preprint page with a generated annotation button.}
  \label{fig:button}
\end{figure*}

\begin{figure*}[ht]
  \centering
  \begin{tikzpicture}
    \node[draw,line width=1pt,inner sep=0pt] (image) {\includegraphics[width=1\textwidth]{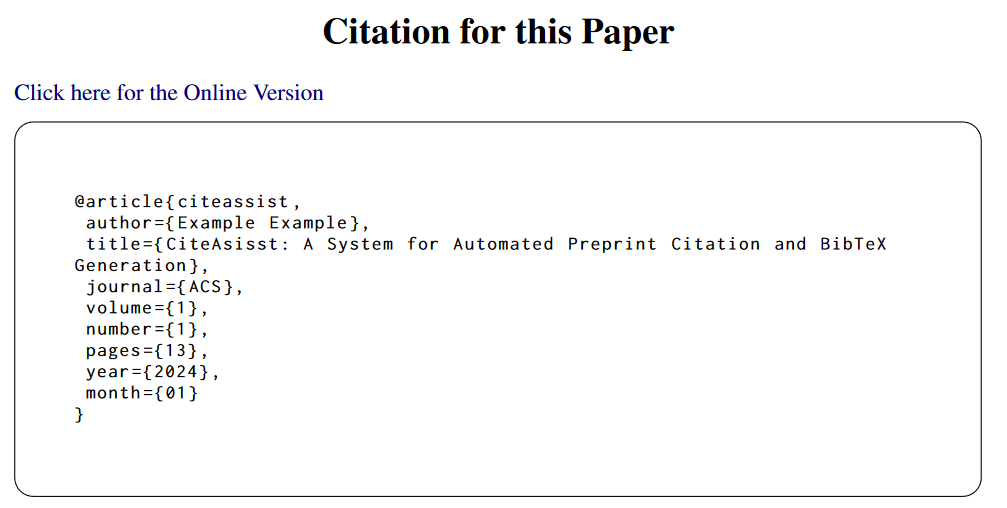}};
  \end{tikzpicture}
  \caption{A generated citation box on the last page of the preprint.}
  \label{fig:annotation-large}
\end{figure*}

\clearpage
\onecolumn
\input{ai-usage-card}
\newpage
\input{annotation}

\end{document}

%% file: ai-usage-card.tex
{\sffamily
    \centering
    \tcbset{colback=white!10!white}
    \begin{tcolorbox}[
        title={\large \textbf{AI Usage Card for \textit{Enhanced Preprint Generator}} \hfill \makebox{\qrcode[height=1cm]{https://ai-cards.org}}},
        breakable,
        boxrule=0.7pt,
        width=.8\paperwidth,
        center,
        skin=bicolor,
        before lower={\footnotesize{AI Usage Card v1.0 \hfill \url{https://ai-cards.org} \hfill \href{https://jpwahle.com/ai-cards-preprint}{PDF} | \href{https://jpwahle.com/cite/jcdl2023wahle.bib}{BibTeX}}},
        segmentation empty,
        halign lower=center,
        collower=white,
        colbacklower=tcbcolframe]
            
        \footnotesize{
            \begin{longtable}{p{.15\paperwidth} p{.275\paperwidth} p{.275\paperwidth}}
              {\color{LightBlue} \MakeUppercase{Correspondence(s)}} \newline Lars Benedikt Kaesberg
              & {\color{LightBlue} \MakeUppercase{Contact(s)}} \newline \href{mailto:l.kaesberg@stud.uni-goettingen.de}{l.kaesberg@stud.uni-goettingen.de}
              & {\color{LightBlue} \MakeUppercase{Affiliation(s)}} \newline Georg-August University Göttingen
              \\\\
              & {\color{LightBlue} \MakeUppercase{Project Name}} \newline CiteAssist 
              & {\color{LightBlue} \MakeUppercase{Key Application(s)}} \newline automate preprint generation
              \\\\
              {\color{LightBlue} \MakeUppercase{Model(s)}} \newline chat-gpt
              & {\color{LightBlue} \MakeUppercase{Date(s) Used}} \newline 2023-06-07
              & {\color{LightBlue} \MakeUppercase{Version(s)}} \newline 3.5\\\\
              \cmidrule{2-3}\\
      
              {\color{LightBlue} \MakeUppercase{Ideation}} \newline    
              & {\color{LightBlue} \MakeUppercase{Generating ideas, outlines, and workflows}} \newline Not used 
              & {\color{LightBlue} \MakeUppercase{Improving existing ideas}} \newline Not used \\\\
              & {\color{LightBlue} \MakeUppercase{Finding gaps or compare aspects of ideas}} \newline Not used \\\\
              
              {\color{LightBlue} \MakeUppercase{Literature Review}} \newline chat-gpt   
              & {\color{LightBlue} \MakeUppercase{Finding literature}} \newline search for similar work/tools
              & {\color{LightBlue} \MakeUppercase{Finding examples from known literature}} \newline Not used \\\\
              & {\color{LightBlue} \MakeUppercase{Adding additional literature for existing statements and facts}} \newline Not used
              & {\color{LightBlue} \MakeUppercase{Comparing literature}} \newline Not used \\\\
              \cmidrule{2-3}\\
      
              {\color{LightBlue} \MakeUppercase{Methodology}} \newline    
              & {\color{LightBlue} \MakeUppercase{Proposing new solutions to problems}} \newline Not used
              & {\color{LightBlue} \MakeUppercase{Finding iterative optimizations}} \newline Not used \\\\
              & {\color{LightBlue} \MakeUppercase{Comparing related solutions}} \newline Not used \\\\
              
              {\color{LightBlue} \MakeUppercase{Experiments}} \newline    
              & {\color{LightBlue} \MakeUppercase{Designing new experiments}} \newline Not used
              & {\color{LightBlue} \MakeUppercase{Editing existing experiments}} \newline Not used \\\\
              & {\color{LightBlue} \MakeUppercase{Finding, comparing, and aggregating results}} \newline Not used \\\\
              \cmidrule{2-3}\\
      
              {\color{LightBlue} \MakeUppercase{Writing}} \newline chat-gpt   
              & {\color{LightBlue} \MakeUppercase{Generating new text based on instructions}} \newline Not used
              & {\color{LightBlue} \MakeUppercase{Assisting in improving own content}} \newline rewrite own text to improve writing style \\\\
              & {\color{LightBlue} \MakeUppercase{Paraphrasing related work}} \newline Not used 
              & {\color{LightBlue} \MakeUppercase{Putting other works in perspective}} \newline Not used \\\\
              
              {\color{LightBlue} \MakeUppercase{Presentation}} \newline    
              & {\color{LightBlue} \MakeUppercase{Generating new artifacts}} \newline Not used
              & {\color{LightBlue} \MakeUppercase{Improving the aesthetics of artifacts}} \newline Not used \\\\
              & {\color{LightBlue} \MakeUppercase{Finding relations between own or related artifacts}} \newline Not used \\\\
              \cmidrule{2-3}\\
              {\color{LightBlue} \MakeUppercase{Coding}} \newline    
              & {\color{LightBlue} \MakeUppercase{Generating new code based on descriptions or existing code}} \newline Not used
              & {\color{LightBlue} \MakeUppercase{Refactoring and optimizing existing code}} \newline Not used \\\\
              & {\color{LightBlue} \MakeUppercase{Comparing aspects of existing code}} \newline Not used \\\\
              
              {\color{LightBlue} \MakeUppercase{Data}} \newline    
              & {\color{LightBlue} \MakeUppercase{Suggesting new sources for data collection}} \newline Not used 
              & {\color{LightBlue} \MakeUppercase{Cleaning, normalizing, or standardizing data}} \newline Not used  \\\\
              & {\color{LightBlue} \MakeUppercase{Finding relations between data and collection methods}} \newline Not used  \\\\
              \cmidrule{2-3}\\
      
              {\color{LightBlue} \MakeUppercase{Ethics}} \newline    
              & {\color{LightBlue} \MakeUppercase{What are the implications of using AI for this project?}} \newline undefined
              & {\color{LightBlue} \MakeUppercase{What steps are we taking to mitigate errors of AI for this project?}} \newline undefined \\\\
              & {\color{LightBlue} \MakeUppercase{What steps are we taking to minimize the chance of harm or inappropriate use of AI for this project?}} \newline undefined
              & {\color{LightBlue} \MakeUppercase{The corresponding authors verify and agree with the modifications or generations of their  used AI-generated content}} \newline Yes \\
      
            \end{longtable}
        }
        \tcblower
    \end{tcolorbox}
}

%% file: annotation.tex
\hypertarget{annotation}{}
\citationtitle

\onlineversion{https://preprint.larskaesberg.de/preprint/54584256-0c0a-4657-8239-166ceb45d736}
\begin{bibtexannotation}
@inproceedings{kaesberg2024,
 author={Lars Kaesberg, Terry Ruas, Jan Philip Wahle, Bela Gipp},
 title={CiteAssist: A System for Automated Preprint Citation and BibTeX Generation},
 pages={15},
 year={2024},
 month={07},
 confacronym={SDProc @ ACL 2024}
}\end{bibtexannotation}

\begin{relatedpapers}
    \relatedpaper{Behera, Prashanta Kumar and Jain, Sanmati Jinendran and Kumar, Ashok. Visual Exploration of Literature Using Connected Papers: A Practical Approach. http://dx.doi.org/10.29173/istl2760. 2023. }
\end{relatedpapers}